# An Enterprise Architecture Framework for E-learning

1. Abbas Najafizadeh, Islamic Azad University, Shahr-e-Qods Branch, Tehran, Iran,
najafizadeh@shahryariau.ac.ir

2. Maryam Saadati, Islamic Azad University, Tehran North Branch, Tehran, Iran,
m_saadati@iau-tnb.ac.ir

3. S. Mahdi Jamei, Islamic Azad University, Shahr-e-Qods Branch, Tehran, Iran,
jamei@shahryariau.ac.ir

4. S. Shervin Ostadzadeh, Science & Research Branch of IAU, Tehran, Iran,
ostadzadeh@srbiau.ac.ir

Abstract

With a trend toward becoming more and more information and communication based, learning services and processes were also evolved. E-learning comprises all forms of electronically supported learning and teaching. The information and communication systems serve as a fundamental role to implement these learning processes. In the typical information-driven organizations, the E-learning is part of a much larger platform for applications and data that extends across the Internet and intranet/extranet. In this respect, E-learning has brought about an inevitable tendency to lunge towards organizing their information based activities in a comprehensive way. Building an Enterprise Architecture (EA) undoubtedly serves as a fundamental concept to accomplish this goal. In this paper, we propose an EA for E-learning information systems. The presented framework helps developers to design and justify completely integrated learning and teaching processes and information systems which results in improved pedagogical success rate.

Introduction

It goes without saying that nowadays utilizing information and communication technologies in enterprises is one of the most challenging tasks. E-learning comprises all forms of electronically supported learning and teaching. The information and communication systems, whether networked or not, serve as specific media to implement the learning process [1]. This is due to the fact that proper development of information and communication systems would have substantial impact on tuning activities in E-learning. The dynamic nature of activities in the E-learning involves changes in corresponding information systems and this imposes considerable maintenance costs to educational enterprises. These costs usually make managers reluctant to commit to regular changes and degrading gradual developments. Enterprise Architecture is the novel promising concept to address this problem, intended to unify an enterprise and the underlying information and communication technologies by employing a structured framework and methodology.

To have a thorough understanding of activities going on in an (educational) enterprise, one should perceive different views of components contained in that enterprise. Each view represents an architecture and refers to one entity. For example, data architecture represents a single view of the data structures used by an enterprise and its applications. They contain descriptions of data in storage and data in motion; descriptions of data stores, data groups and data items; and mappings of those data artifacts to data qualities, applications, locations, etc. Enterprise Architecture refers to a collection of architectures which are assembled to form a comprehensive view of an enterprise. Organizing such great amounts of information requires a framework. To address this problem, various enterprise architecture frameworks have emerged [2], such as, FEAF [3], TEAF [4], TOGAF [5] and C4ISR [6]. Zachman Framework (ZF) [7,8] originally proposed by John Zachman, is often referenced as a standard approach for expressing the basic elements of enterprise architecture.

In this paper we propose a framework for E-learning inspired by Zachman Framework. ZF is widely accepted as the main framework in EA. Compared to other proposed frameworks, it has evident advantages to list [9]: (1) using well-defined perspectives; (2) using comprehensive abstracts; (3) normality; and (4) extensive usage in practice. They were the motivations for ZF adoption in our work.

## Methods

Before we discuss our methods, let's briefly introduce some basic concepts and principles. We believe these remarks can help readers to clearly understand what we mean by the concepts that are used in this article.

**Architecture**

Architecture has emerged as a crucial part of design process [9]. Generically, architecture is the description of a set of components and the relationships between them. According to ISO 15704 [10], an architecture is a description of the basic arrangement and connectivity of parts of a system (either a physical or a conceptual object or entity). In computer science, there are software architectures, hardware architectures, network architectures, information architectures, and enterprise (IT) architectures.

**Enterprise**

According to ISO 15704, an enterprise is one or more organizations sharing a definite mission, goals and objectives to offer an output such as a product or a service [10]. An enterprise consists of people, information, and technologies; performs business functions; has a defined organizational structure that is commonly distributed in multiple locations; responds to internal and external events; has a purpose for its activities; provides specific services and products to its customers [9].

**Enterprise Architecture (EA)**

Generally speaking, an Enterprise Architecture should be organized in a way that supports reasoning about the structure, properties and behavior of the system [11]. It defines the components that make up the overall system and provides a blueprint from which the system can be developed. EA shows the primary components of an enterprise and depicts how these components interact with or relate to each other [9]. EA typically encompasses an overview of the entire information system in an enterprise; including the software, hardware, and information architectures.

**Zachman Framework (ZF)**

In 1987, an IBM researcher, named John A. Zachman, proposed a framework for Information System Architecture, which is now called Zachman Framework [7]. ZF is a two dimensional information matrix consisting of 6 rows and 6 columns.

The rows describe the perspectives of various stakeholders. These rows starting from the top include: Planner (Scope), Owner (Enterprise Model), Designer (System Model), Builder (Technology Model), Contractor (Detail Representation), and Functioning Enterprise.

The columns describe various abstractions that define each perspective. These abstractions are based on six questions that one usually asks when s/he wants to understand a thing. The columns include: Data (What is it made of?), Function (How does it work?), Network (Where are the elements?), People (Who does what work?), Time (When do things happen?), and Motivation (Why do things happen?). To find cell definitions of ZF refer to [12].

**A ZF-based Method for E-learning**

As mentioned earlier, we introduce our proposed framework based on ZF intended to support all aspects of the E-learning information system as a whole. Similar to ZF, it is a two dimensions matrix. The columns are the same as Zachman Framework. They are based on six basic interrogatives that are asked to understand an entity. The rows represent various perspectives of the pedagogical approaches for E-learning.

Results

Figure 1 depicts the proposed E-learning framework schema. As you can see, it is a matrix with 4 rows and 6 columns. The columns include:

- Data (What is it made of?): This focuses on the material composition of the information. In the case of E-learning, it focuses on data that can be used for pedagogical systems.

| E-learning Framework | Data | Function | Network | People | Time | Motivation |
|---|---|---|---|---|---|---|
| *Cognitive* | | | | | | |
| *Contextual* | | | | | | |
| *Service* | | | | | | |
| *Technical* | | | | | | |

Fig. 1: A framework for E-learning based on Zachman Framework

- Function (How does it work?): This focuses on the functions or transformations of the information used for pedagogical systems.

- Network (Where are the elements?): This focuses on the geometry or connectivity of the data used for pedagogical systems.

- People (Who does what work?): This focuses on the people and the manuals and the operating instructions or models used for pedagogical systems.

- Time (When do things happen?): This focuses on the life cycles, timings and schedules used for pedagogical systems.

- Motivation (Why do things happen?): This focuses on goals, plans and rules that prescribe policies and ends that guide pedagogical systems.

The rows, starting from the top, are:

- Cognitive: This is the highest level of abstraction in the framework. It refers to the cognitive boundaries of the pedagogy from the E-learning systems view. This row focuses on the cognitive processes involved in learning.

- Contextual: This row stands below the cognitive layer, and focus on the contextual and environmental aspects which can stimulate E-learning. It refers to the skills and behavioural outcomes of the interaction with others for learning process.

- Service: This row stands beneath the Contextual layer. In this row, the services that the information systems should provide in the E-learning will be addressed. This is the perspective of a pedagogical expertise.

- Technical: This is the lowest level of abstraction in the framework. It focuses on basic technologies that the E-learning system should provide to support service layer. This row is the perspective of the E-learning system development team.

## Discussions

The framework contains 24 cells. Each cell describes a model that one might document to describe the E-learning system. Each of the cells in the framework is primitive and thus, each can be described or modeled independently. All of the cells on a given row make up a given perspective. All of the cells in a column are related to each other since they focus on the same type of elements. Following sections present the framework's cells definitions.

**Data Column**

Cognitive-Data cell is simply a list of important pedagogical entities (or objects, or assets) that the E-learning is interested in. It is probably adequate that this list is at a fairly high level of aggregation. It defines the scope, or boundaries, of the rows 2-4 models of entities.

Contextual-Data cell is a model of the actual pedagogical entities (objects, assets) that are significant to the E-learning system. It typically would be at a level of definition that it would express concepts (terms and facts) used in the significant pedagogical objectives/strategies that would later be implemented.

Service-Data cell is a model of the pedagogical entities (objects, assets) that are significant to the data of the E-learning services.

Technical-Data cell would be the definition of all the data specified by the Service-Data Model and would include all the data definition language required for the E-learning system implementation.

**Function Column**

Cognitive-Function cell is simply a list of important pedagogical processes (or functions) that the E-learning system performs. It is probably adequate that this list is at a fairly high level of aggregation. It defines the scope, or boundaries, of the rows 2-4 models of processes.

Contextual-Function cell is a model of the actual pedagogical processes (or functions) that the E-learning performs. It can be represented as a model expressing the pedagogical transformations (processes) and their inputs and outputs.

Service-Function cell is a model of the pedagogical processes (or functions) that are significant to support the function of the E-learning services.

Technical-Function cell would be the algorithms specifications for the E-learning system implementation.

**Network Column**

Cognitive-Network cell is simply a list of locations in which the E-learning operates. It is probably adequate that this list is at a fairly high level of aggregation. It defines the scope, or boundaries, of the models of locations that are connected by the E-learning and are found in rows 2-4.

Contextual-Network cell is a model of the actual pedagogical locations (or nodes) and their connections that are significant to the E-learning system. It would include identification of the types of facilities at the nodes and connections.

Service-Network cell is a model of the pedagogical locations (or nodes) and connections that are significant to the locations of the E-learning services.

Technical-Network cell is the specific definition of the node addresses and the lines required for the E-learning system implementation.

**People Column**

Cognitive-People cell is simply a list of important pedagogical organizations to which the E-learning assigns responsibility for work. It is probably adequate that this list is at a fairly high level of aggregation. It defines the scope, or boundaries, of the models that are responsible to the E-learning and depicted on rows 2-4.

Contextual-People cell is a model of the actual pedagogical allocation of responsibilities and specification of E-learning system.

Service-People cell is a model of the pedagogical responsibilities that are significant to the work of the E-learning services.

Technical-People cell is the physical expression of work flow of the E-learning including the specific individual and their requirements and the presentation format of the work product required for E-learning system implementation.

**Time Column**

Cognitive-Time cell is simply a list of important pedagogical events to which the E-learning responds. It is probably adequate that this list is at a fairly high level of aggregation. It defines the scope, or boundaries, of the time models that are significant to the E-learning and found in Rows 2-4.

Contextual-Time cell is a model of the actual pedagogical time that is comprised of an initiating event and an elapsed time. It typically would be at a level of definition that it would express sequence, or relative time.

Service-Time cell is a model of the pedagogical time (events) that are significant to the E-learning services. This model describes the E-learning events that trigger the state to transition from one valid state (point in time) to another and the dynamics of that transition cycle.

Technical-Time cell is the definition of time and events required for E-learning system implementation.

**Motivation Column**

Cognitive-Motivation cell is simply a list of major pedagogical goals (or objectives, or strategies, or critical success factors) that are significant to the E-learning. It is probably adequate that this list is at a fairly high level of aggregation. It defines the scope, or boundaries, of the models of goals (etc.) that are embraced by the E-learning and found in the constructs of Rows 2-4.

Contextual-Motivation cell is a model of the pedagogical objectives and strategies (the "ends" and "means") that constitute the motivation behind the E-learning system.

Service-Motivation cell is a model of the pedagogical rules of the E-learning services in terms of their intent ("ends") and the constraints ("means").

Technical-Motivation cell would be the specification of the pedagogical rules required for the E-learning system implementation.

## Conclusion

In this paper, a framework for E-learning based on Zachman Framework is proposed. Compared to other frameworks, ZF has some evident advantages. These advantages have caused its extensive usage as the basic framework of enterprise architecture, and are the motivations for ZF adoption in our work.

Education & IT integration is a critical challenge faced by E-learning industry. The proposed framework can be an initial step to demonstrate how the models in the E-learning system can be made coherent in order to avoid inconsistencies within the framework. The framework helps developers to design and justify completely integrated education, technologies, and information systems which results in improved pedagogical success rate.